\documentclass[twocolumn,showpacs,preprintnumbers,amsmath,amssymb,groupedaddress,superscriptaddress,floatfix]{revtex4}

\newif\ifpdf
\ifx\pdfoutput\undefined
\pdffalse 
\else
\pdfoutput=1 
\pdftrue
\fi
\ifpdf
\usepackage[pdftex]{graphicx}
\usepackage[pdftex]{color}
\DeclareGraphicsExtensions{.jpg,.png,.pdf}
\else
\usepackage[dvips]{graphicx}
\usepackage[dvips]{color}
\DeclareGraphicsExtensions{.eps}
\fi

\usepackage{dcolumn}
\usepackage{bm}
\newcommand{\ignore}[1]{}

\begin{document}


\title{Diffusion-induced Ramsey  narrowing}
\author{Yanhong Xiao}
\author{Irina Novikova}
\author{David F. Phillips}
    \affiliation{Harvard-Smithsonian Center for Astrophysics,
Cambridge, Massachusetts, 02138}
\author{Ronald L. Walsworth}
\affiliation{Harvard-Smithsonian Center for Astrophysics,
Cambridge, Massachusetts, 02138} \affiliation{Department of
Physics, Harvard University, Cambridge, Massachusetts, 02138}

\date{\today}

\begin{abstract}
     A novel form of Ramsey narrowing is identified and characterized.
For long-lived coherent atomic states coupled by laser fields, the diffusion
of atoms in-and-out of the laser beam induces a spectral narrowing of
the atomic resonance lineshape.  Illustrative experiments and an
intuitive analytical model are presented for this diffusion-induced
Ramsey narrowing, which occurs commonly in optically-interrogated
systems.

\end{abstract}


%
%

\maketitle


The lifetime of an atomic coherence is often limited by the finite
interaction time between the atoms and resonant radiation: e.g., by
atomic motion through a laser beam. For atoms constrained to diffuse
in a buffer gas, this interaction time is usually estimated by the
lowest order diffusion mode, which leads to the typical Lorentzian
lineshape, but implicitly assumes that atoms diffuse out of the laser
beam and do not return~\cite{happer72,arimondo'96pra, Erhard}.
However, when other decoherence effects are small, atoms can diffuse
out of the interaction region and return before decohering.  That is,
atoms can evolve coherently in the dark (outside of the laser beam)
between periods of interaction (inside the laser beam), in analogy to
Ramsey spectroscopy~\cite{Ramsey}. In many cases of interest,
diffusing atoms can spend a majority of their coherence lifetime in the
dark, which induces a significant spectral narrowing of the center of
the atomic
resonance lineshape.

\begin{figure}
\includegraphics[width=7.5cm]{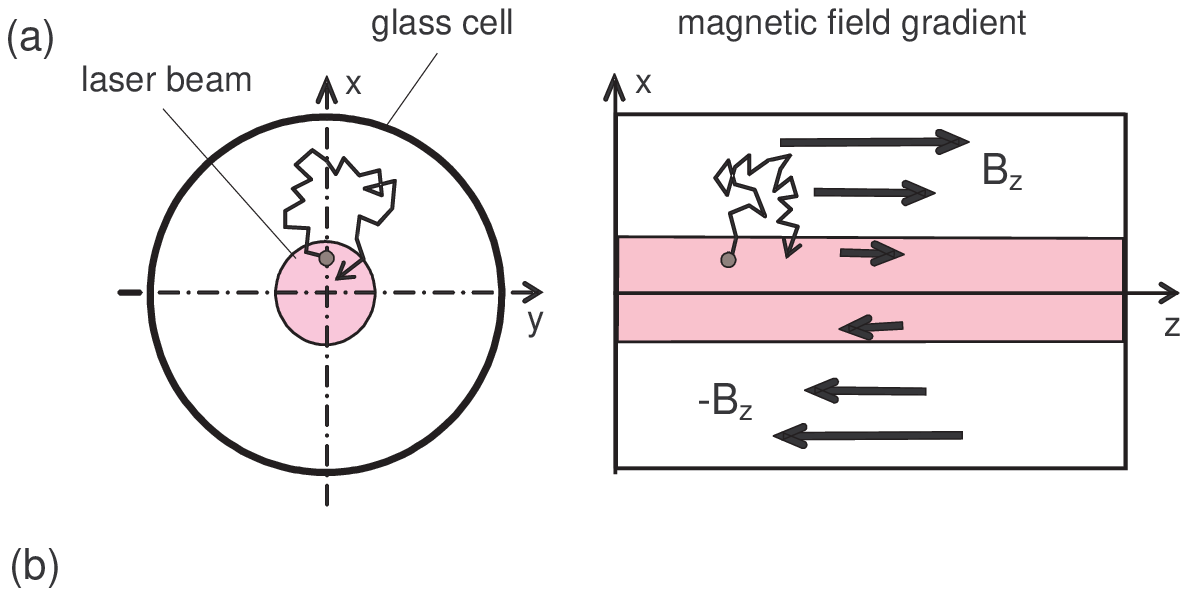}
\includegraphics[width=7.5cm]{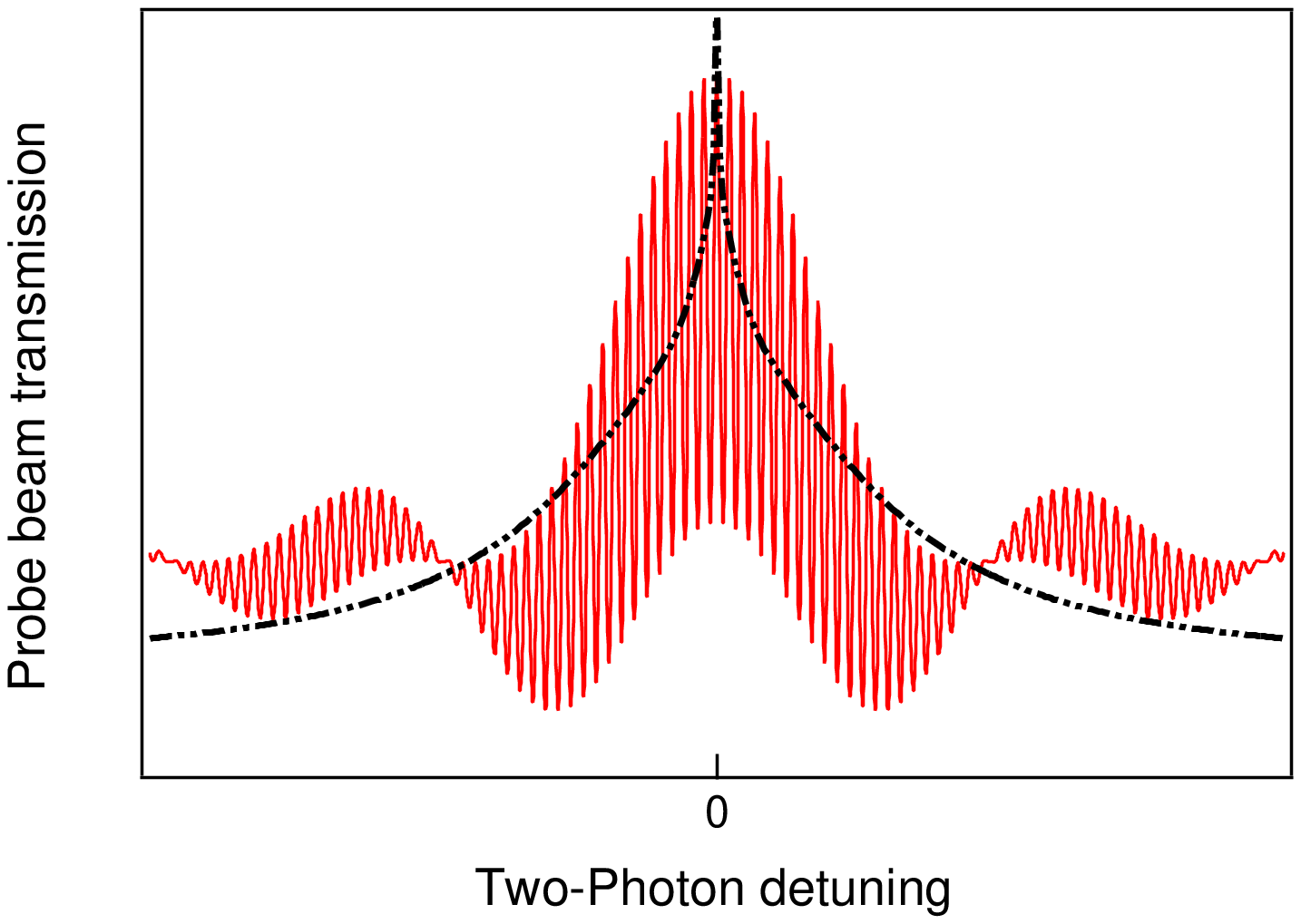}
\caption{ (a) \textit{Left}: an example path for atoms
diffusing in and out of the laser beam. \textit{Right}: in some of
the measurements reported here, a transverse gradient in the
longitudinal magnetic field was applied to the atomic vapor cell in
order to decohere most atoms that diffused out of the laser beam. (b)
Calculated EIT lineshapes for repeated diffusive
return of atomic coherence to the laser beam. Solid red curve: an
example lineshape for a particular diffusion history (``Ramsey
sequence''), with the time spent in the laser beam $t_{in}=\tau_D$
(the mean diffusion time to leave the beam given
by the lowest-order diffusion mode), and the time spent out of the
beam before returning $t_{out}= 20\tau_D$. Dashed black curve: weighted
average over all Ramsey sequences. } \label{ramsey.fig}
\end{figure}

In the present Letter, we identify this ``diffusion-induced Ramsey
narrowing'' as a general phenomenon, which
we characterize through demonstration experiments using
Electromagnetically Induced Transparency (EIT) in warm Rb vapor, and
with an intuitive analytical model of the repeated diffusive return
of atomic coherence to the laser beam (see Fig.~\ref{ramsey.fig}).
The effects identified here are particularly important for atomic
frequency standards~\cite{vanier05apb} and for dynamic light-matter
interactions such as slow and stored light in atomic
vapor~\cite{lukin03rmp}, but to date have only been treated in a few
special cases~\cite{Zibrov-optics,Zibrov-PRA}.

EIT results from optical pumping of atoms into a noninteracting
``dark'' state for two optical fields that are in two-photon Raman
resonance with a pair of metastable ground states of the atomic
system~\cite{arimondo'96,scullybook}. EIT gives rise to a narrow
transmission resonance for the optical fields, with a minimum
spectral width set by the rate of decoherence between the two ground
states constituting the ``dark'' state. To characterize
diffusion-induced Ramsey narrowing using EIT, we employed a diode
laser operating at 795 nm on the ${}^{87}$Rb $D1$ transition to drive the
atoms into EIT resonance between the $F= 2$ and $F= 1$ hyperfine
levels of the electronic ground state. The beam passed through an
enriched ${}^{87}$Rb vapor cell (2.5 cm diameter, 5 cm length, Ne
buffer gas) which was heated to approximately
45\,${}^\circ$C to create optically thin Rb vapor ($n\sim 6\times10^{10}$
cm$^{-3}$). The cell was mounted within three layers of magnetic
shields to screen external fields. Sets of coils were used as needed
to provide a homogeneous longitudinal magnetic field ($B_z$) and/or
a transverse gradient in the longitudinal field ($\partial
B_z/\partial x$) as shown in Fig.~\ref{ramsey.fig}a. A
photodetector measured the total light intensity transmitted through the vapor
cell.

\begin{figure}
\includegraphics[width=7.5cm]{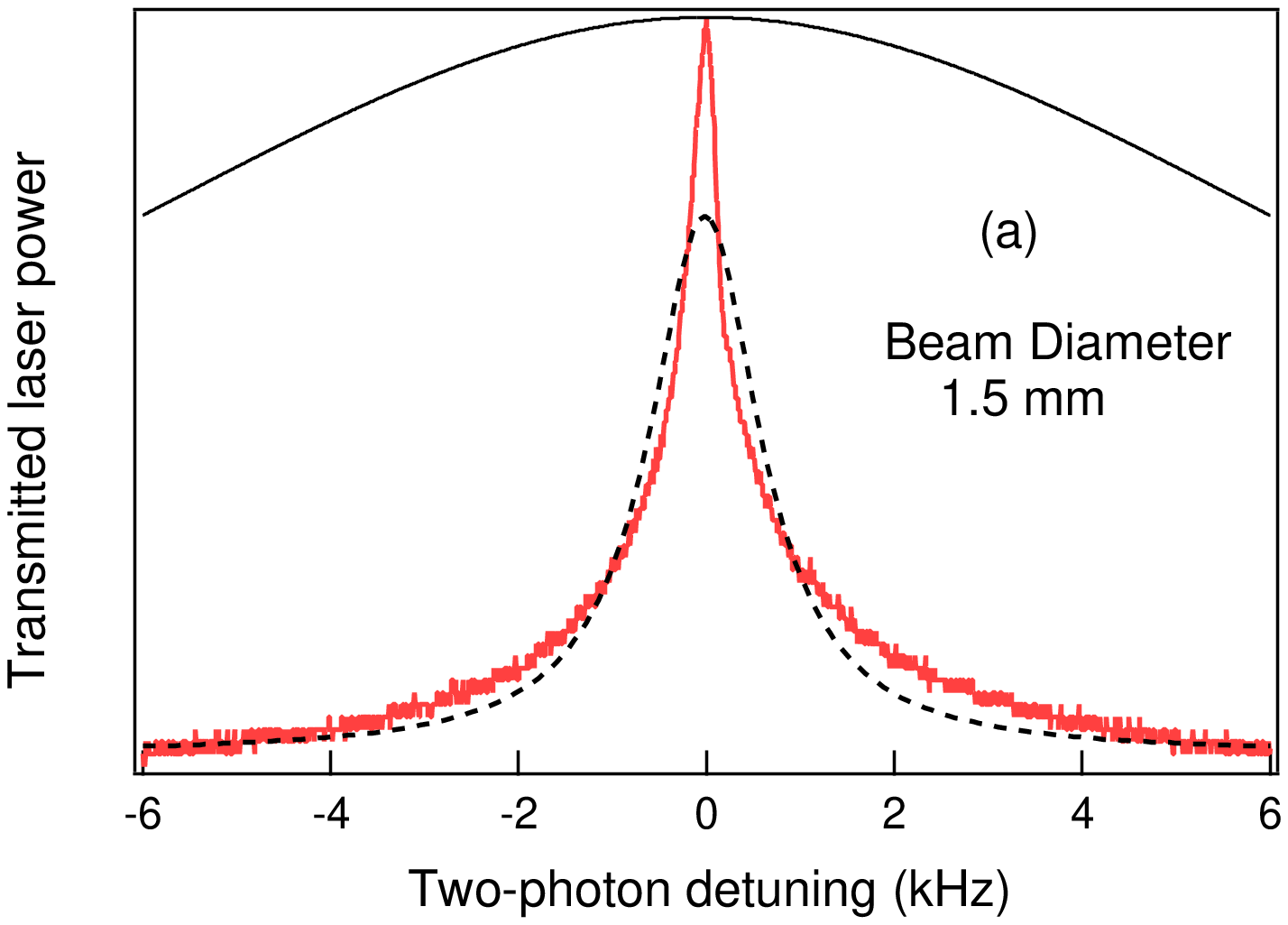}
\includegraphics[width=7.5cm]{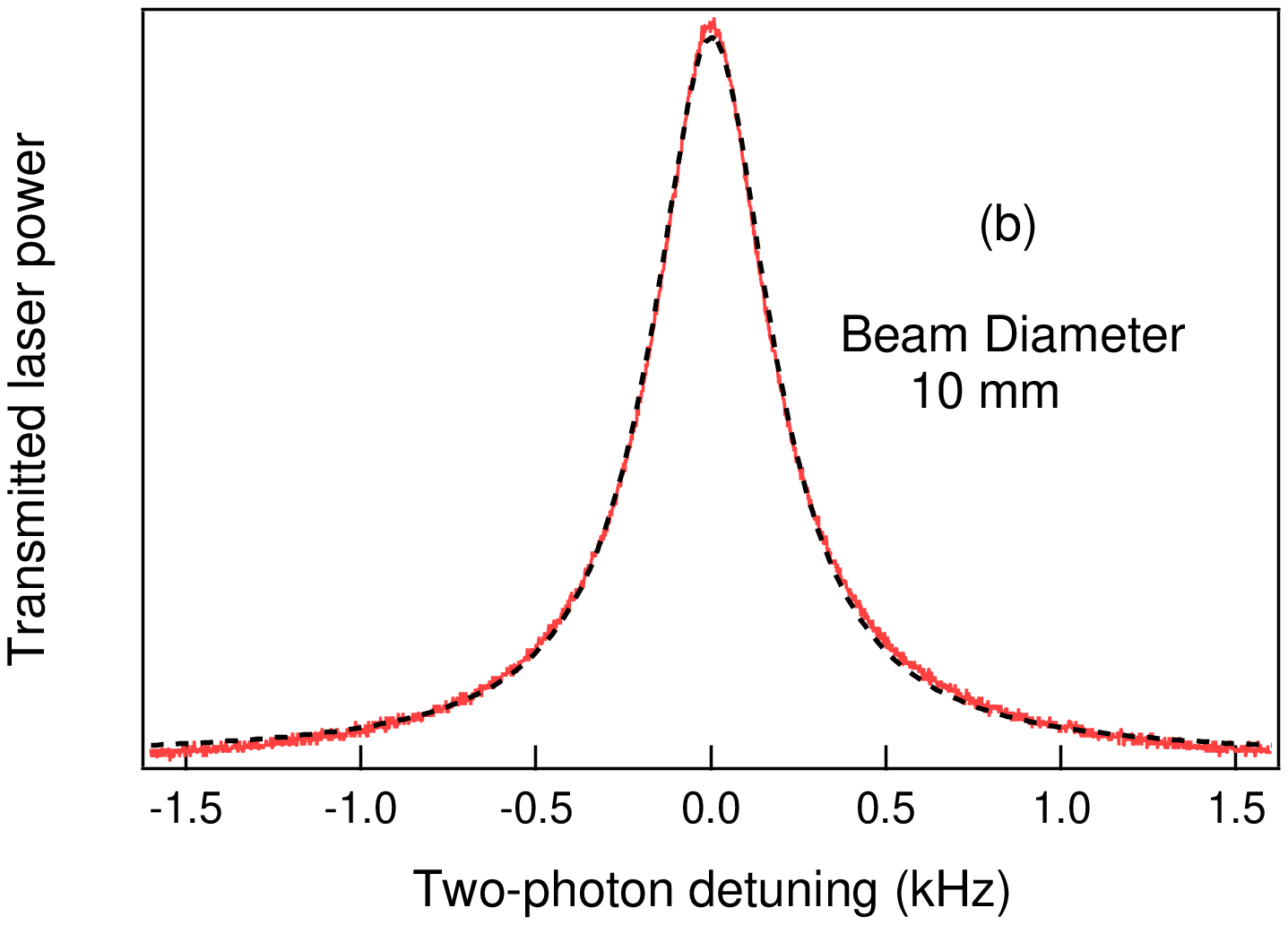}
\caption{Measured Rb EIT lineshapes (in red) with laser beam
   diameters of (a) 1.5 mm and (b) 10 mm, in a 3 Torr Ne cell with 70
   $\mu$W laser power in the two EIT fields.  Dashed lines are fits to
   a Lorentzian lineshape. Broad solid curve in (a) is a 20 kHz FWHM
   Lorentzian, the expected lineshape for a coherence lifetime set by
   the lowest order diffusion mode out of the beam.}
\label{EITbeam.fig}
\end{figure}

In a first set of experiments, we employed a VCSEL (vertical cavity
surface emission laser) with a transverse Gaussian intensity profile,
which was current-modulated at 3.4~GHz to form the two optical fields
necessary for EIT ~\cite{VCSEL}. We used a collimating lens and iris,
and measured EIT lineshapes for various laser beam diameters in a
cell with 3 Torr Ne buffer gas (Rb diffusion coefficient $D\approx$
50 cm$^2$s$^{-1}$). We set $B_z$ = 0 and $\partial
B_z/\partial x$ = 0, such that EIT occurred for all
relevant combinations of $m_F$ sublevels. As shown in
Fig.~\ref{EITbeam.fig}a, the measured EIT resonance for a 1.5 mm
diameter beam has a full-width-half-maximum (FWHM) of 740
Hz, whereas the calculated FWHM $\approx$ 20 kHz if one makes the 
common assumption that the coherence lifetime is set by the lowest 
order diffusion mode out of the beam (e.g., see~\cite{Erhard}). As 
also shown in Fig.~\ref{EITbeam.fig}a, the measured EIT lineshape for 
a 1.5 mm diameter beam is spectrally narrower near-resonance than a 
Lorentzian: this sharp central peak is the characteristic signature 
of diffusion-induced Ramsey narrowing. In contrast, the measured EIT
resonance for a 10 mm diameter beam is well fit by a Lorentzian 
lineshape with FWHM $\approx$ 400 Hz (see Fig.~\ref{EITbeam.fig}b), 
which is in good agreement with the calculated FWHM using the lowest 
order diffusion mode, and is consistent with the
small fraction of atoms that leave this relatively large diameter
beam and return during the maximum coherence lifetime (set by buffer
gas collisions and diffusion to the cell walls).

\begin{figure}
\includegraphics[width=7.5cm]{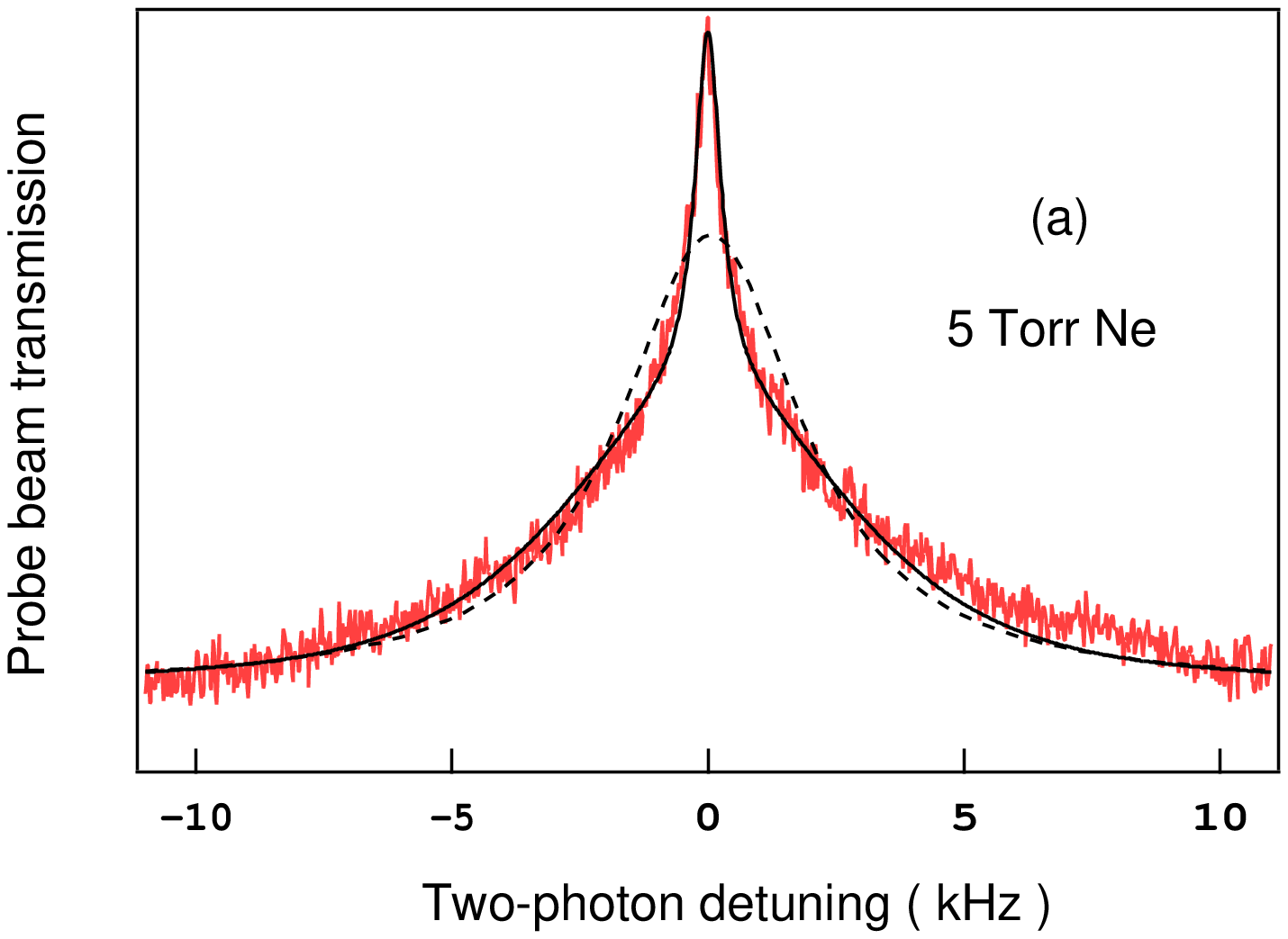}
\includegraphics[width=7.5cm]{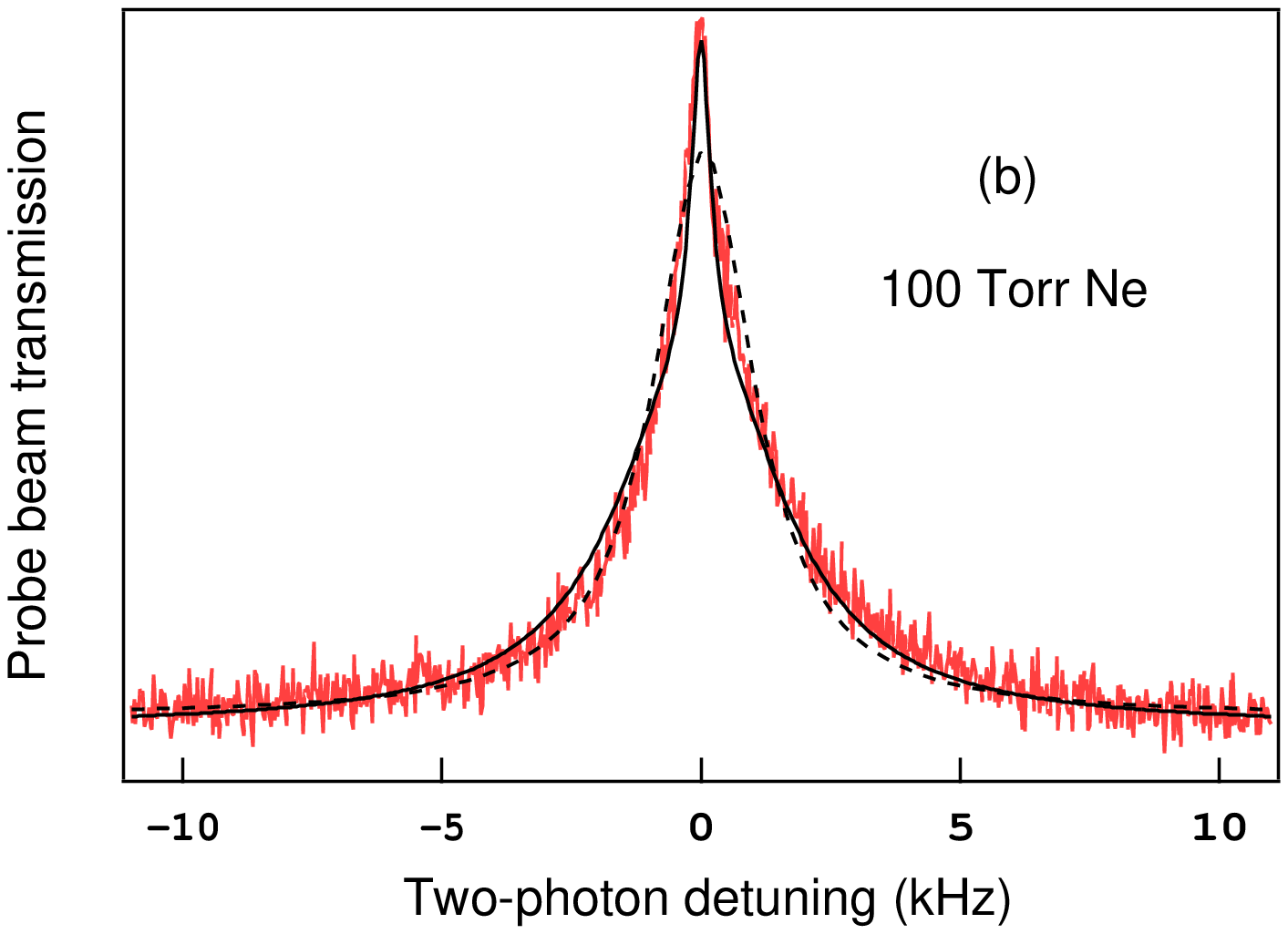}
\caption{Measured Rb EIT lineshapes (in red) at (a) 5 Torr and (b)
   100 Torr Ne buffer gas pressure, with $22~\mu$W laser power and 0.8
   mm beam diameter. Dashed lines are fits to a Lorentzian lineshape.
   Solid lines are results from the repeated interaction model. }
\label{EITpressure.fig}
\end{figure}

In a complementary set of experiments, we measured the EIT lineshape
as a function of buffer gas pressure, thereby altering the Rb
diffusion coefficient and changing the fraction of atomic coherence
that evolves in the dark. (We used a slightly different apparatus
than that described above. See~\cite{JMO} for details.) In
Fig.~\ref{EITpressure.fig} we compare measured EIT lineshapes for 5
Torr and
100 Torr Ne buffer gas ($D\approx$ 30  and 1.5 cm$^2$s$^{-1}$,
respectively), with a 0.8 mm laser beam diameter. Fits to the data
are shown both for our analytical ``repeated interaction model''
(outlined below) and a Lorentzian lineshape. The repeated
interaction model provides a good fit at both high and low buffer gas
pressure, demonstrating its ability to account for the physics of
diffusion-induced Ramsey narrowing.
In contrast, the Lorentzian fit deviates significantly from the
measurements even for 100 Torr Ne buffer gas --- i.e., even under
conditions of relatively slow Rb diffusion and reduced coherence
evolution in the dark --- which indicates the inadequacy of the
traditional approach of assuming that atoms diffuse out of the laser beam
and do not return.

\begin{figure}
\includegraphics[width=7.5cm]{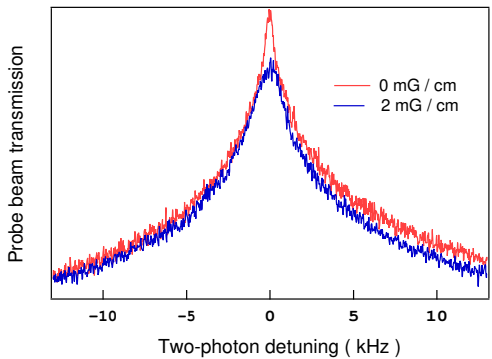}
\caption{Measured modification of the Rb EIT lineshape by application of a
     transverse magnetic field gradient $\partial B_z/\partial x$ = 2
mG/cm in a 5 Torr Ne
     buffer gas cell with 0.8 mm laser beam diameter.  The field
gradient suppresses the sharp peak caused by diffusion-induced Ramsey
narrowing, but leaves the broader EIT lineshape largely unaffected.}
     \label{EITgradient.fig}
\end{figure}

Destroying the coherence of atoms that leave the laser beam should
eliminate diffusion-induced Ramsey narrowing. As a demonstration of
this behavior, we measured magnetic-field-sensitive EIT spectra
(coupling the $F=2$, $m_F=1$ and $F=1$, $m_F=1$ levels) in the
presence of both a longitudinal magnetic field $B_z \approx 43$ mG and
a transverse gradient in the longitudinal magnetic field, $\partial
B_z/\partial x$ (Fig.~\ref{ramsey.fig}a), using a 0.8 mm diameter
laser beam and 5 Torr Ne buffer gas~\cite{JMO}. An appropriately
chosen magnetic field gradient ($\approx$ 2 mG/cm) induced modest
inhomogeneous broadening of the EIT resonance ($\approx$ 400 Hz, smaller than 
the width of the sharp central peak) for
atoms within the total volume of the laser beam~\cite{JMO}, but caused most atoms that diffused out
of the beam to lose phase coherence because of the extended period
such atoms typically spend in the dark (hence probing larger
differences in the magnetic field). As shown in
Fig.~\ref{EITgradient.fig}, the characteristic sharp peak in the EIT
spectrum is effectively eliminated by application of the magnetic
field gradient.

\begin{figure}
\includegraphics[width=7.5cm]{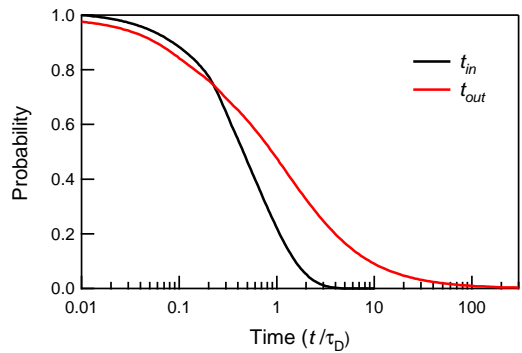}
\caption{Probability distributions, calculated from the diffusion
equation, for an atom to spend time $t_{in}$ inside a laser beam with
a step-like transverse intensity profile, and then to spend time
$t_{out}$ outside of the laser beam before returning. Times are
expressed relative to $\tau_D$, the mean diffusion time to leave the
beam given by the lowest-order diffusion mode. } \label{t_distr.fig}
\end{figure}

We found good agreement between our measurements and numerical
calculations of the Maxwell-Bloch equations, which describe the
atom-light interaction, coupled to the diffusion equation, which
describes the atomic motion. We also developed and successfully
applied a more intuitive and analytically-soluble approach --- the
repeated interaction model mentioned above. In this model, the
atomic resonance lineshape is calculated for an atom having a
specific history (``Ramsey sequence") of alternating interactions
with the laser beam and evolution in the dark. The lineshape for the
atomic ensemble is then determined by a weighted average of the
lineshapes from different Ramsey sequences,
using the distributions of times spent in and out of the laser beam
($t_{in}$ and $t_{out}$) as
determined from the diffusion equation; see
Fig.~\ref{t_distr.fig}. With this approach, the atomic motion and the atomic
response to laser fields are decoupled, which dramatically simplifies
the calculation and allows for an analytical solution.

For example, Fig.~\ref{ramsey.fig}b shows the EIT lineshape calculated
analytically for one particular Ramsey sequence, as well as the
ensemble lineshape determined from the weighted average over Ramsey
sequences. Under the assumptions that an atom spends equal time
$t_{in}$ in the laser beam before and after diffusing in the dark for
a period $t_{out}$, as well as a large
difference in intensity between the two EIT optical fields, the
analytical expression for the weak EIT field's transmission $\mathrm{T}$
as a function of two-photon Raman detuning $\Delta$ is given by:
\begin{eqnarray}\label{long-equation}
   \mathrm{T}(\Delta)&=&\mathrm{T}_0+\frac{\kappa |\Omega_d|^2 \eta}{\Delta
     ^2+\Gamma^2}\left(-\Gamma+ \sqrt{\Delta ^2+\Gamma^2} \ \times\right.
   \\ \nonumber && \left\{ e^{-\Gamma t_{in}} \cos \left[\Delta\cdot
       t_{in}+\phi_\vartriangle\right]- \right.
   \\ \nonumber
   &&\, \, \, e^{-\Gamma t_{in}-\Gamma_0 t_{out} } \cos \left[\Delta\cdot
     (t_{out}+t_{in})+\phi_\vartriangle\right]+
   \\ \nonumber
   &&\left.\left. \, \, e^{-2\Gamma t_{in} -\Gamma_0 t_{out} } \cos
       \left[\Delta\cdot
         (2t_{in}+t_{out})+\phi_\vartriangle\right] \right\}\right).
\end{eqnarray}
Here $\mathrm{T}_0$ is the background transmission far from two-photon
Raman resonance through the optically thin cell;
$\kappa=\frac{3\pi}{16}n\lambda^2L/\gamma^2$, where $n$ is the atomic
density, $\lambda$ is the optical wavelength, $L$ is the cell length,
and $\gamma$ is the relaxation rate of the excited state; $\Omega_d$
is the Rabi frequency for the strong optical field; $\eta$ is the radiative 
decay rate of the excited state;
$\Gamma=\Gamma_0+|\Omega_d|^2/2\gamma$ is the power-broadened EIT
linewidth in the absence of diffusion-induced Ramsey narrowing, where
$\Gamma_0$ is the intrinsic relaxation rate of the ground-state
coherence (set by buffer gas collisions, etc.); and
$\tan{\phi_\vartriangle}=\Delta/\Gamma$. The above expression also
assumes $\gamma\gg\Delta, \Gamma_0, \Gamma$ and
$\gamma\Gamma\gg\Delta^2$, which are typically satisfied for EIT in
warm Rb vapor. The first term in brackets in Eq.~(\ref{long-equation})
is the contribution from atoms that interact with the laser beam only
once. The second and third terms account for returning atoms. More
generally, the time spent inside the laser beam before and after
leaving may differ. Also, atoms may return to the beam more than once;
each additional diffusive return will produce two extra terms similar
to the last two lines in Eq.~(\ref{long-equation}).

To achieve good fits to the measured EIT lineshapes with no free
fitting parameters (see Fig.~\ref{EITpressure.fig}), we found that it
is sufficient to consider Ramsey sequences limited to only one or two
evolution periods in the dark.  In these calculations we assumed a
step-like laser profile in the transverse direction, which is a good
approximation when the effective two photon Rabi period is longer
than the average time atoms spend in the laser beam, so that an atom
averages over the transverse Gaussian distribution of laser
intensity. Details of these calculations are described
in~\cite{diffModel}.

The width of the lineshape envelope for an individual Ramsey
sequence, such as that shown in Fig.~\ref{ramsey.fig}b, scales
inversely with the time the atom
spends in the laser beam $(t_{in}\approx\tau_D \propto w^2/D$, where $D$ is
the Rb diffusion coefficient and $w$ is the beam width); whereas the
width of the Ramsey fringes is inversely proportional to the free
evolution time in the dark $(t_{out} > \tau_D)$. The sharp central
peak, indicative of
diffusion-induced Ramsey narrowing, emerges intuitively in this
model, since only the central fringe adds constructively for all Ramsey
sequences, with different diffusion times outside of the beam.  The
narrow width of this central peak is limited by other effects (atomic
collisions, magnetic field gradients, wall collisions,
etc.) which set an upper bound on the free evolution time. Since the
atoms contributing most to the sharp central peak spend the majority of
their time in the dark, the width of this peak is relatively
insensitive to power broadening.

In general, when the laser beam diameter is small, reshaping and
narrowing of the lineshape are strong, since a large fraction of the
atoms participate in the diffusion-induced Ramsey process, and the
free-evolution time between interactions with the laser beam can be
long. For larger
laser beam diameters, the Ramsey narrowing gradually
disappears since a smaller fraction of the atoms can diffuse out of
the beam and return before
decohering due to other effects.  In particular, when the laser beam
diameter approaches
the cell diameter, atoms diffusing out of the beam rapidly decohere
due to wall collisions. (In ongoing work we are investigating the
effect of coherence-preserving wall-coatings.)

We note that non-Lorentzian EIT lineshapes can appear in other
circumstances.  For instance, a well-known form of linewidth
narrowing occurs in optically thick media due to frequency-selective
absorption~\cite{density-narrowing}. Alternatively, for an optically
thin medium with inhomogeneous laser intensity, atoms can reach
equilibrium locally in the limit of high buffer gas pressure,
producing a spatial
variation of the power broadening and an inhomogeneously broadened (and
non-Lorentzian) lineshape~\cite{TY, Levi}; whereas in the limit of low buffer
gas pressure, atoms can be pumped at one intensity and probed at
another, leading to a
non-Lorentzian lineshape dependent on the effusive time-of-flight of
atoms across the sample cell \cite{time-of-flight}. These effects are
qualitatively and quantitatively distinct from diffusion-induced
Ramsey narrowing.

In conclusion, we identified a novel form of spectral narrowing
arising from the diffusion of atomic coherence in-and-out of an 
optical interaction
region, such as a laser beam. We characterized this
``diffusion-induced Ramsey narrowing" with measurements on
Electromagnetically Induced Transparency (EIT) in warm Rb vapor, and
found good agreement with an intuitive analytical model based on a
weighted average of distinct atomic histories in the light and the
dark. This ``repeated interaction model'' and the spectral
narrowing effects studied here are
relevant to spectroscopy, quantum optics
and other applications based on long-lived atomic coherences.

We are grateful to F. Can\`e, J. Vanier and A.\ B.\ Matsko for useful
discussions, and to C. Smallwood and C. Wang for assistance in
experiments and modelling. This work was supported
by ONR, DARPA, and the Smithsonian Institution.

\end{document}